\documentclass[lettersize,journal]{IEEEtran}

\usepackage{setspace}

\usepackage{amsmath,amsfonts}
\usepackage{algorithm}
\usepackage{array}
\usepackage[caption=false,font=normalsize,labelfont=sf,textfont=sf]{subfig}
\usepackage{textcomp}
\usepackage{stfloats}
\usepackage{url}
\usepackage{verbatim}
\usepackage{graphicx}
\usepackage{cite}
\usepackage{bm}
\usepackage{pgfplots}

\usepackage{algorithm}
\usepackage{algpseudocode}
\usepackage{listings}
\usepackage{xcolor} 
\definecolor{mygreen}{rgb}{0,0.6,0}
\usepackage{array}  
\usepackage{paralist}
\usepackage{colortbl}
\usepackage{amssymb}
\usepackage{pifont}

\usepackage{amssymb}  
\usepackage{tikz}
\usepackage{diagbox}  
\usepackage{adjustbox} 
\usepackage{verbatim}
\usepackage{caption}
\usepackage{rotating}  
\usepackage{multirow}
\usepackage{makecell}  
\usepackage{arydshln}

\title{\textsc{Thor}: A  Non-Speculative Value Dependent Timing Side Channel Attack Exploiting Intel AMX}
\author{Farshad Dizani, Azam Ghanbari, Joshua Kalyanapu, Darsh Asher, and Samira Mirbagher Ajorpaz}

\lstdefinestyle{boxed}{
    frame=single, 
    xleftmargin=3pt,
    xrightmargin=3pt,
    aboveskip=4pt,
    belowskip=0pt,
}

\lstdefinestyle{nonboxed}{
    frame=none, 
    xleftmargin=0pt, 
    xrightmargin=0pt,
    aboveskip=4pt,
    belowskip=0pt,
}

\usepackage{graphicx}

\newcolumntype{P}{>{\hspace{0.3cm}}p{0.9cm}}
\newcolumntype{C}{>{\centering\arraybackslash}m{3cm}}  
\newcolumntype{X}{>{\centering\arraybackslash}m{4cm}}  

\newcommand{\paragrabf}[1]{\noindent \textbf{#1}\ }

\usepackage{titlesec}
\titlespacing{\section}{0pt}{*0.8}{*0.5}
\titlespacing{\subsection}{0pt}{*0.5}{*0.3} 

\begin{document}





 \maketitle

\begin{spacing}{0.96}
    
\begin{abstract}
The rise of on-chip accelerators signifies a major shift in computing, driven by the growing demands of artificial intelligence (AI) and specialized applications. These accelerators have gained popularity due to their ability to substantially boost performance, cut energy usage, lower total cost of ownership (TCO), and promote sustainability. Intel's Advanced Matrix Extensions (AMX) is one such on-chip accelerator, specifically designed for handling tasks involving large matrix multiplications commonly used in machine learning (ML) models, image processing, and other computational-heavy operations.

In this paper, we introduce a novel value-dependent timing side-channel vulnerability in Intel AMX. By exploiting this weakness, we demonstrate a software-based, value-dependent timing side-channel attack capable of inferring the sparsity of neural network weights without requiring any knowledge of the confidence score, privileged access or physical proximity. Our attack method can fully recover the sparsity of weights assigned to 64 input elements within 50 minutes, which is 631\% faster than the maximum leakage rate achieved in the Hertzbleed attack.
\end{abstract}


\section{Introduction}
Privacy of machine learning models deployed on Machine Learning as a Service (MLaaS) platforms can be compromised by 
adversaries to extract sensitive details, including model architecture
and hyperparameters~\cite{TramerZJRR16}. 
However, this paper shows that 
the tight integration of on-core AI accelerators like Intel AMX\cite{intel2023amx} causes AMX performance to depend upon operand value which works as proxy, enabling timer attacks on ML models thus introducing a timing side channel that bypasses conventional defenses like speculative execution mitigations 
or cache isolation 
and challenges the architectural and data privacy of ML.

Cache attacks exploit timing variations between accessing or flushing data in cache levels or DRAM~\cite{GrussFlushFlush}.
Spectre~\cite{kocher2020spectre} and Meltdown~\cite{meltdown} exploit speculative execution and exception handling. 
MDS Attacks~\cite{schwarz2019zombieload,canella2019fallout} target microarchitectural buffers. Defenses include cache partitioning~\cite{kiriansky18dawg}, 
Invisispec~\cite{yan2018invisispec}, privileged access controls (e.g., RAPL restrictions)~\cite{Lipp2021Platypus}, SMT disabling and other patches with performance overhead. 
%
However, no work has been done on the security evaluation of newly built Intel AMX for AI applications.


Black-box attacks extract model functionality or hyperparameters with minimal access but require confidence scores and are mitigated by  various methods like limiting the query rate, obfuscating the input data by masking, rounding confidence, or using homomorphic encryption.  
Early physical attacks, such as IKWYS~\cite{wei2018know}  
used power or EM side channels to infer neural network (NN) parameters but required physical access. 
~\cite{Yan2020Cachetelepathy} 
leverages cache timing to leak NN architecture but is reliant on particular ML libraries and co-location. 
 We confirmed that vulnerabilities on NNs using floating-point timing~\cite{9218707} have been completely mitigated in the Intel AMX design and thus are no longer available to the attacker. 

 In this paper, we demonstrate the feasibility of exploiting Intel AMX 
 to leak sensitive information in NN, such as sparsity of it's weights. Inferring sparsity could inadvertently reveal sensitive data patterns. Understanding the sparsity pattern might provide insights into the underlying data distributions, potentially breaching data privacy.
Additionally, knowledge of sparsity patterns could enable attackers to craft more effective adversarial attacks. By understanding the structure of a model, an attacker might design inputs that strategically exploit sparsity, leading to degraded model performance or incorrect predictions.
 Our approach is notable for not requiring physical access or a shared cache, unlike previous works 
 that often necessitating physical access for power and electromagnetic measurements 
 or shared cache manipulation via known cache side channels \cite{Yan2020Cachetelepathy}, lowering the bar for potential attackers.
We discover that the execution times of the Intel AMX multiplications vary depending on the operand's value. 
This phenomenon is different from the observed Hertzbleed\cite{wang2022hertzbleed, liu2022frequency}, which is a side-channel attack that leverages dynamic voltage and frequency scaling (DVFS) on modern x86 processors to transform power fluctuations into timing attacks, extractable without direct power measurement. Hertzbleed exploits CPU frequency variations, linked to data-dependent power consumption. Our attack continues to function even with the Intel Turbo Boost (DVFS) feature turned off, suggesting a novel root cause specific to the Intel AMX design compared to Hertzbleed, which relies on the DVFS feature being enabled and obsolete cryptographic libraries.
Our attack also bypasses TEE based defenses in contrary to Rowhammer-based attacks like DeepSteal~\cite{RakinDeepSteal2021} that leak partial weights but can be protected by TEE-based defenses. 


%
We successfully leaked zeros in 64 hidden weights with 100\% accuracy within 50 minutes of observation without access to the NN output value, confidence score or shared cache bypassing state of the art defenses such as obfuscation, masking or rounding confidence values of NN and cache defenses.
Our proof of concept (PoC) attack  (\textsc{Thor}) represents the first step toward a new direction in ML privacy attacks. It is an attack on a single-layer NN but its significance is that, unlike prior approaches that rely on {{confidence scores}}, logits,  physical access, or higher privilege, \textsc{Thor} leverages timing information alone to infer critical and private parameters like NN's weight sparsity patterns and distributions. 

This paper presents the following contributions:
\begin{compactitem}
\item It characterized and reverse-engineered the Intel AMX, 
uncovering 
a novel type of data-dependent timing side-channel vulnerability within the Intel AMX accelerator.

\item It demonstrates the techniques necessary to exploit this newly discovered timing side-channel for launching attacks on NNs. Specifically, it illustrates how to determine the sparsity of NN weights by measuring execution time.
We show that the \textsc{Thor} leakage rate is 631\% and 1,493\% higher than the maximum leakage rates in the Hertzbleed and Collide+Power~\cite{kogler2023collide+} attacks, respectively.
\item We discuss defenses, propose a counter-measurement and measure its overhead on power consumption (2.59\% to 12.33\%).   
\end{compactitem}
\vspace{0.1em}
\textbf{Responsible disclosure.}  We reported our attack and findings to Intel in May 2024 and Intel confirmed our findings. 

\section{Threat Model}
\begin{figure}[!ht] 
    \centering
   \includegraphics[width=\linewidth]{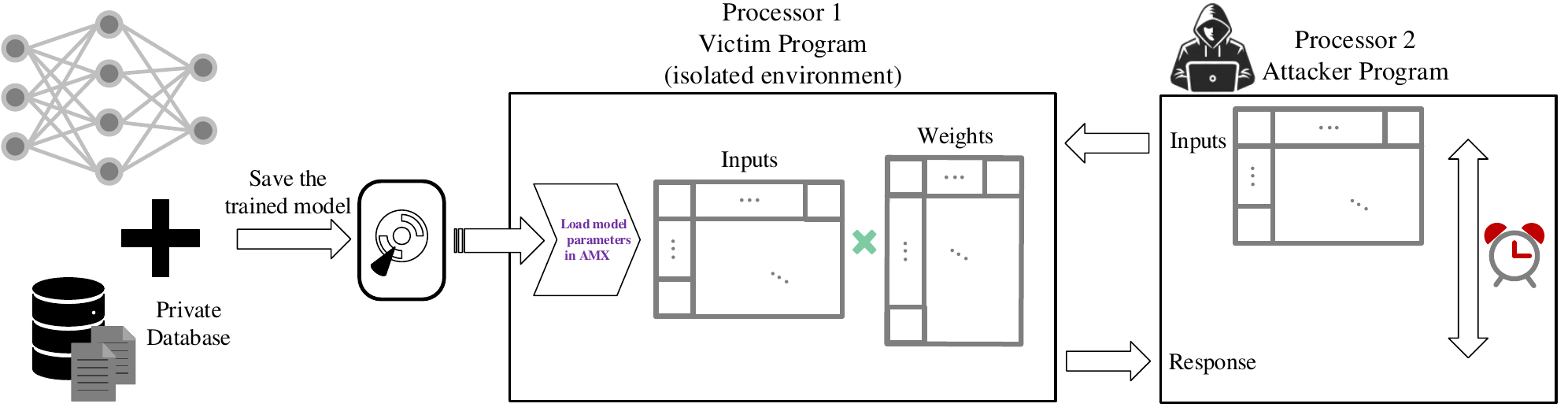} 
    \caption{\textsc{Thor} Threat Model.}
    \label{fig:Thor-threat-model}
\end{figure}

The attacker and victim are two distinct processes running on the same server. These processes are entirely isolated in terms of address space and can be executed on different CPU cores. As shown in Figure \ref{fig:Thor-threat-model}, the victim process is an API that runs a NN designed for specific tasks and utilizes Intel AMX for matrix computations. The attacker process, which operates as a non-privileged user, does not have direct access to the model’s parameters but interacts with the model via the inference API. The attacker sends crafted inputs to the inference API and measures the timing of the victim process's responses. By examining timing variations, the attacker can gain insights into the NN's internal parameters by taking advantage of the timing differences caused by Intel AMX computations. We assume cache defenses are deployed or there is no shared cache,  thus channels such as \cite{Yan2020Cachetelepathy} is not available to the adversary. The ML model is a single layer NN protected by obfuscated the confidence score and the output by methods such as adding noise or rounding confidence scores \cite{Fredrikson2015ModelInversion}. The victim may leverage TEEs, Intel SGX, to execute the NN computations within isolated enclaves.


\vspace{-1em}\section{Thor}
Our investigations utilized a server running {Red Hat Enterprise Linux release 9.4} with {Linux kernel version 5.14}, powered by an { Intel Xeon Gold 5420+} from the {Sapphire Rapids microarchitecture}. 
The tests were carried out with a range of configurations,
including the { autonomous (hardware-only)} and { cooperative P-state control modes}, under {minimal power}, {maximum performance}, and {efficiency-focused} platform power settings. {Turbo Boost} was tested in both {enabled} and {disabled states}. {Prefetchers} were toggled between {enabled} and {disabled states} across experiments to assess their impact on noise and timing variations. The server configurations included the UEFI firmware versions SRV650-v3-3.14 (May 2024) and SRV650-v3-3.20 (June 2024), with the latter mitigating certain Spectre variants but all above settings remains vulnerable to our attack. 

\subsection{Reverse Engineering}
\begin{figure}[htbp!]
    \centering
    \begin{tikzpicture}
        \begin{axis}[
            font=\scriptsize,
            width=\linewidth*0.95,
            height=\linewidth*0.35,
            bar width=50,
            xmin=35500, xmax=56000,
            xlabel={Execution time (cycle)},
            ybar,
            ylabel={Frequency},
            ylabel style={yshift=-15pt},
            ymin=0, ymax=0.0024,
            legend style={
                at={(0.5,1)}, 
                anchor=north, 
                legend columns=-1,
                font=\scriptsize,
                /tikz/column 2/.style={column sep=5pt}, 
                fill=none,
                draw=none
            },
            legend image code/.code={\draw[fill=#1,draw=none] (0cm,-0.1cm) rectangle (0.25cm,0.1cm);},
            legend entries={100\% sparsity, 50\% sparsity, 0\% sparsity}
        ]
            \addplot+[hist={density, bins=60}, fill=lightgray, draw=none] table[y index=0] {Diagrams/Data/1000-0.txt};
            \addplot+[hist={density, bins=60}, fill=black, draw=none] table[y index=0] {Diagrams/Data/1000-half.txt};
            \addplot+[hist={density, bins=60}, fill=gray, draw=none] table[y index=0] {Diagrams/Data/1000-1.txt};
        \end{axis}
    \end{tikzpicture}
\caption{Execution Time with Varying Operand Sparsity. }

\label{fig:Cumulative_execution_time}
\end{figure}
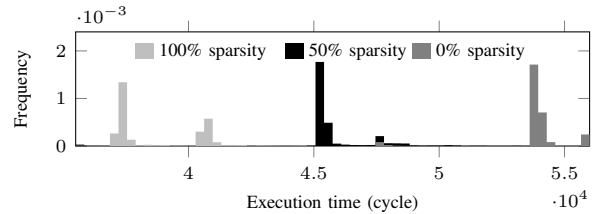

Intel AMX is an on-core accelerator introduced with the 4th Gen Intel Xeon Scalable processors, specifically engineered to boost matrix computation efficiency.
%
Tile Matrix Multiply unit (TMUL) is the accelerator engine for executing multiplication calculations. 
Intel AMX can perform 512 and 1024 multiplications/additions per cycle for BF16 and INT8, respectively~\cite{IntelOptRefManual}. 
We studied the impact of operand sparsity on TMUL operations. 
By varying the number of zero values in the Tile operands during multiplication, we observed that increasing the number of zeros accelerates the 
AMX execution. 
This effect is illustrated in Figure~\ref{fig:Cumulative_execution_time}
, which presents three distinct distributions showing the sparsity levels of the operand: 100\% sparsity, 50\% sparsity, and 0\% sparsity. 
The results indicate that a greater sparsity of operands in the tile operands leads to faster 
execution of the TMUL. The program with 1,000 AMX multiplications averages 54,005 cycles when operands have 0\% sparsity. Increasing sparsity to 50\% and 100\% reduces the average execution time to 45,953 and 38,747 cycles, respectively. In this case, 50\% and 100\% more sparsity make the execution 17.5\% and 39.4\% faster, respectively.
%

\begin{figure}[h!]
\centering
\vspace{-0.5cm}\begin{tikzpicture}
    \begin{axis}[
        xlabel={Number of Instructions},
        ylabel={Execution Time (RDTSCP cycle)},
        legend pos=outer north east,
        font=\scriptsize,
        xlabel style={yshift=5pt},
        ylabel style={yshift=-15pt},
        width=\linewidth*1,
        height=\linewidth*0.4,
        xmin= -0.1,xmax=40000,
        legend style={
                legend columns=-1,
                at={(0.5,1.16)},
                anchor=north, 
                font=\tiny,
                /tikz/column 2/.style={column sep=5pt}, 
            },
    ]
    
    \addplot[
        line width=0.8pt,
        color=black,
    ] table {Diagrams/Data/1400MHz.txt};
    \addlegendentry{1.2 GHz
CPU frequency};

    \addplot[
        line width=0.8pt,
        color=lightgray,
    ] table {Diagrams/Data/2000MHz.txt};
    \addlegendentry{2 GHz
CPU frequency};

    \end{axis}
    
    \node[darkgray, thick] at (2.7,1.3) {\scriptsize 1 GHz};
    \node[darkgray, thick] at (5.7,0.97) {\scriptsize 1.2 GHz};
    \node[darkgray, thick] at (2.5,0.83) {\scriptsize 1.3 GHz};
    \node[darkgray, thick] at (5.5,0.32) {\scriptsize 2 GHz};
\end{tikzpicture}
\caption{Illustration of AMX Frequency Adjustment}
\label{fig:Freq_transition}
\vspace{-1em}\end{figure}
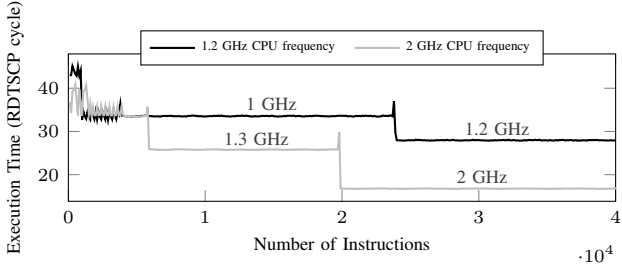

This discovery uncovers a new value-dependent timing side channel that reveals the sparsity of Tile operands in Intel AMX. Our results suggest the presence of a zero-skipping-like mechanism for both BF16 and INT8 operands in Intel AMX design for acceleration and power saving purposes. 
Since the operand value timing dependence persists even with Turbo Boost disabled and the CPU frequency fixed, we hypothesized that AMX operates under a separate clock domain. To estimate its internal independent frequency (assuming there is one), we analyzed its performance and observed that the execution times initially fluctuate through intermediate levels before stabilizing. We measured the stabilized execution time of AMX multiplications at each available CPU frequency. By comparing these times, we found that interestingly the intermediate levels correspond precisely to some frequencies in our dataset. This suggests that AMX transitions through intermediate frequencies before stabilizing to match the CPU frequency. For example, at a fixed CPU frequency of 1.2 GHz, AMX stabilizes at 1 GHz before aligning with 1.2 GHz. Similarly, at a fixed CPU frequency of 2 GHz, AMX progresses through 1 GHz and 1.3 GHz before reaching 2 GHz (see Figure~\ref{fig:Freq_transition}). These consistent transitions across tests strongly suggest that AMX may operate under an independent clock, autonomously adjusting its frequency to match the CPU frequency. Interestingly, the latency of this adjustment also depends on operand values.

\subsection{Attack Building Blocks}
The \textsc{Thor} attack begins with strategically crafted input vectors, where the adversary measures \texttt{TMUL} execution time under different operand conditions. Multiple measurements mitigate inherent noise, yielding a composite timing value. 

\paragrabf{(1) Setting Threshold.}
By comparing execution times with inputs of full zeros and full non-zeros, we establish a differential threshold (\(Thr\)). This threshold enables the selective acceptance of randomly generated input vectors for scoring. The essence of the attack involves two key steps: (1) continuously generating random input vectors throughout the attack and (2) using the execution time associated with each vector to accumulate evidence that helps refine the estimates of the weight scores.
Suppose a weight element is non-zero; then, the execution time will be longer if the corresponding input element is non-zero compared to when it is zero. Conversely, if a weight element is zero, the execution time remains unchanged whether the corresponding input element is zero or not. Based on these assumptions, if a generated input vector has a higher number of correct non-zeros according to the unknown weights, it is more likely to be accepted by this algorithm, thereby impacting the scoring tables.

\begin{figure}[!htbp]
\centerline{\includegraphics[width=0.65\linewidth]{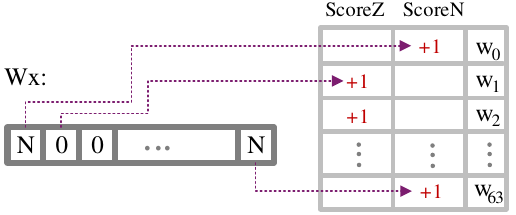}}
\caption{Scoring Update Process Diagram: \(ScoreZ\) and \(ScoreN\) increment based on \(W_x\) values.}
\label{fig:ML_Scoring}
\end{figure}

\paragrabf{(2) Execution Time Analysis and Vector Selection.}
The attacker measures execution times for two scenarios: one for the randomly generated vector (\(W_s\)) and another for its inverted version (\(W_r\)) (in \(W_r\), all zeros in \(W_s\) become non-zeros, and all non-zeros become zeros). We predict that the vector holding more correct non-zeros according to the unknown weights will require a longer execution time. If the difference in execution times exceeds the threshold, either \(W_s\) or \(W_r\)—whichever records the longer time—will be selected (\(W_x\)) for updating score vectors (\(ScoreZ\) and \(ScoreN\)). After selecting \(W_x\), a scoring method updates the score vectors according to \(W_x\) entries. This involves iterating through each element of \(W_x\) from 0 to 63, incrementing the corresponding entry in \(ScoreZ\) by one if the element is zero and in \(ScoreN\) if the element is non-zero (see Figure~\ref{fig:ML_Scoring}).

\paragrabf{(3) Inferring Weights from Scoring Vectors.}
Following the attack period, the attacker analyzes the ScoreZ and ScoreN vectors
. A weighted element marked as zero would be expected to yield similar scores in both \(ScoreZ\) and \(ScoreN\) vectors. However, for a weight element marked as non-zero, a higher score is expected in the \(ScoreN\) vector compared to the \(ScoreZ\).
By dividing each element of \(ScoreN\) by \(ScoreZ\), the attacker can infer the zero weights. If the ratio is close to 1, it suggests that the weight scores are equivalent; thus, the weight is likely zero. A ratio significantly above 1 indicates that the weight is likely non-zero.

\section{Security Evaluation}
\subsection{Leakage Rate}
\begin{figure}[htbp!]
\centering
\begin{tikzpicture}
    \begin{axis}[
        xlabel={Attack Duration (minute)},
        ylabel={Success Rate},
        legend pos=outer north east,
        font=\scriptsize,
        xlabel style={yshift=5pt},
        ylabel style={yshift=-15pt},
        width=\linewidth*0.85,
        height=\linewidth*0.3,
        xmax=100, xmin=0,
        legend style={
                legend columns=-1,
                at={(0.5,1.16)},
                anchor=north, 
                font=\tiny,
                /tikz/column 2/.style={column sep=5pt}, 
            },
    ]
    \addplot[
        line width=2pt,
        color=darkgray,
    ] table {Diagrams/Data/accuracy.txt};

    \end{axis}
\end{tikzpicture}
\caption{Impact of Attack Duration on the Success Rate of Weight Determination.}
\label{fig:ML_SuccessRate}
\end{figure}
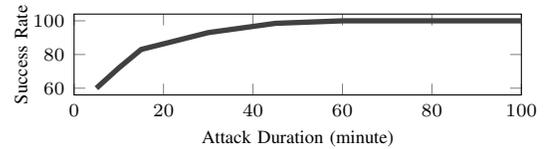

Figure \ref{fig:ML_SuccessRate} shows success rates increasing with longer attack durations, from 60\% at 5 minutes to a complete 100\% success rate at 50 minutes (0.02 bit/s) and beyond.

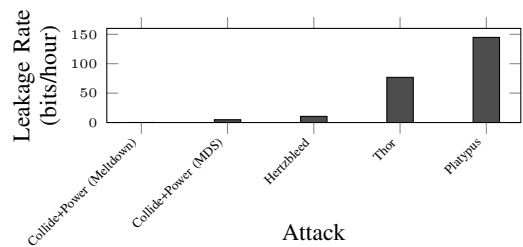
\begin{figure}[!htbp]

    \centering
\vspace{-0.1em}\begin{tikzpicture}
    \begin{axis}[
        ybar,
        symbolic x coords={Collide+Power (Meltdown), Collide+Power (MDS), Hertzbleed, Thor, Platypus},
        xtick=data,
        width=\linewidth*0.8,
        height=\linewidth*0.32,
        xlabel={Attack},
        ylabel={Leakage Rate\\(bits/hour)},
        ymin=0,
        ymax=160,
        x tick label style={
            rotate=45, 
            anchor=east,
            font=\tiny 
        },
        y tick label style={
            font=\tiny 
        },
        bar width=10pt,
        xlabel style={yshift=-0.7cm, font=\small},
        ylabel style={yshift=-0.3cm, align=center, font=\small},
    ]
        \addplot[
        color=black, 
        fill=black!70] coordinates {
            (Collide+Power (Meltdown), 0.136)
            (Collide+Power (MDS), 4.82)
            (Hertzbleed, 10.5)
            (Thor, 76.8)
            (Platypus, 144.7)
        };
    \end{axis}
\end{tikzpicture}
    \caption{\textsc{Thor} leakage rate comparison.}
    \label{fig:thorleakage}
\end{figure}

Figure \ref{fig:thorleakage} shows \textsc{Thor} leakage rate of 76.8 bits/hour compared to other related side channels. While Platypus has a higher leakage rate, it is patched and unavailable to non-privileged users. While \textsc{Thor} does not require access to RAPL and is not limited to SGX, it exhibits a higher leakage rate compared to Hertzbleed and Collide+Power attacks.  
\textsc{Thor}'s leakage rate is {1,493\% faster than Collide+Power (MDS)} (76.8 vs. 4.82 bits/hour), and {631\% faster than Hertzbleed} (10.5 bits/hour), which exploits outdated libraries now replaced by more secure cryptographic implementations. Collide+Power attacks, which leveraged {Microarchitectural Data Sampling (MDS)} and {Meltdown}, have been patched on modern Intel Xeon servers. Similarly, Platypus, which depends on the RAPL interface, is mitigated by restricting access to energy and power reporting. With these mitigations, \textsc{Thor} remains one of the few viable microarchitectural attack methods available to low-privilege attackers, posing a unique and unpatched threat to AI applications, even in environments where traditional microarchitecture side channels have been neutralized.

\subsection{Counter-measurement}
\textsc{Thor} relies only on precise \textit{timing} measurements. Thus, it cannot be mitigated by defenses that alter NN classification outputs, such as adding noise or rounding confidence scores \cite{Fredrikson2015ModelInversion}. Trusted Execution Environment (TEE)-based defenses, like those performing non-linear computations inside the TEE while offloading linear computations to untrusted sources \cite{tramèr2019slalomfastverifiableprivate}, are also inadequate, as the value dependencies of Intel AMX timing that \textsc{Thor} relies on are unaffected by TEE environments like Intel SGX. AI workloads demand high speed and efficiency, prompting AI libraries to prioritize performance optimizations. As a result, known constant-time programming techniques are often unsuitable as a defense. For example, masking can help protect weights from leaking but introduces extra computational overhead, increasing both power consumption and execution time.

A more effective strategy involves introducing response randomness by delaying execution, which disrupts the timing signals relied upon by \textsc{Thor}. However, this approach adds latency to AI applications. Limiting the query rate of the model 
could also slow these attacks, but at the cost of reducing system responsiveness.
Extending detection mechanisms, 
to identify malicious patterns in power and performance characteristics offers a promising defense against \textsc{Thor}.
Lastly, employing homomorphic encryption 
could provide robust protection but comes with substantial computational overhead and performance costs, making it less practical for high-speed AI applications.

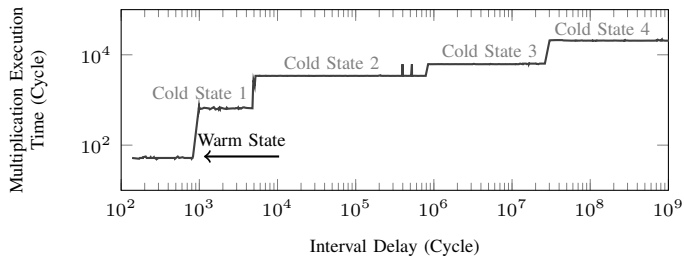
\begin{figure}
    \centering
    \begin{tikzpicture}
        \begin{axis}[
            xmin=100, xmax=1000000000,
            ymin=10, ymax=100000,
            xlabel={Interval Delay (Cycle)},
            ylabel={Multiplication Execution\\Time (Cycle)},
            xmode=log, 
            ymode=log,
            log basis x=10, 
            log basis y=10,
            grid=both,
            major grid style={line width=0pt,draw=white!50},
            minor grid style={line width=0pt,draw=white!20},
            font=\scriptsize,
            ylabel style={yshift=0pt, align=center}, 
            width=\linewidth*1,
            height=\linewidth*0.45,
        ]
            \addplot[darkgray, thick] table {Diagrams/Data/warmup.txt};
        \end{axis}
        
        \draw[->, black, thick] (2.1,0.45) -- (1.1,0.45) node[midway, above] {\scriptsize Warm State};
         \node[gray, thick] at (1.05,1.3) {\scriptsize Cold State 1};
         \node[gray, thick] at (2.8,1.7) {\scriptsize Cold State 2};
         \node[gray, thick] at (4.9,1.85) {\scriptsize Cold State 3};
         \node[gray, thick] at (6.4,2.15) {\scriptsize Cold State 4};
         
    \end{tikzpicture}
    \caption{  Performance States of
TMUL and the Secure Recommended Intel AMX Operational State (Warm).}
    \label{fig:Performance_Stages}
\end{figure}

One mitigation which can be applied through a micro-code update or a software patch is to keep the AMX unit moderately in the Warm State at all times or at least during Intel SGX execution to protect TEEs computation against \textsc{Thor}.
This approach is effective because we observed that timing differences dependent on zero values are only significantly measurable when the Intel AMX is in a Cold State. Warm and Cold States introduced here come from an interesting observation in which 
we measured the time to execute a single AMX multiplication instruction while varying the intervals between consecutive executions. By adjusting the length of these intervals, we identified five distinct execution times, classifying them into performance states. The shortest execution time with the lowest intervals was labeled as the Warm State, while the longer execution times with higher intervals were classified as Cold States.  These performance states are shown in Figure~\ref{fig:Performance_Stages}.
However, this mitigation comes with trade-offs in power management and execution speed, as system power limits could be more easily reached, leading to unnecessary throttling of the AMX unit. We measured the power overhead of such defense for \textsc{Thor} and found that depending on the cold vs. warm stage,  the overhead ranges from 2.59\% to 12.33\%. Although this secure design requires more power consumption, it is faster as it keeps the Intel AMX in the highest performance state at all time; this is in contrary to other secure designs for different microarchitectural attacks which almost always incurred a high performance overhead. 

Thus, future research must prioritize the development of effective mechanisms to mitigate the proposed threat vector introduced by AMX and similar technologies while addressing the secure design's performance and power consumption. 





\end{spacing}

\begin{spacing}{0.95}
    
\bibliographystyle{IEEEtranS}
\bibliography{refs}

\begin{thebibliography}{10}
\providecommand{\url}[1]{#1}
\csname url@samestyle\endcsname
\providecommand{\newblock}{\relax}
\providecommand{\bibinfo}[2]{#2}
\providecommand{\BIBentrySTDinterwordspacing}{\spaceskip=0pt\relax}
\providecommand{\BIBentryALTinterwordstretchfactor}{4}
\providecommand{\BIBentryALTinterwordspacing}{\spaceskip=\fontdimen2\font plus
\BIBentryALTinterwordstretchfactor\fontdimen3\font minus \fontdimen4\font\relax}
\providecommand{\BIBforeignlanguage}[2]{{%
\expandafter\ifx\csname l@#1\endcsname\relax
\typeout{** WARNING: IEEEtranS.bst: No hyphenation pattern has been}%
\typeout{** loaded for the language `#1'. Using the pattern for}%
\typeout{** the default language instead.}%
\else
\language=\csname l@#1\endcsname
\fi
#2}}
\providecommand{\BIBdecl}{\relax}
\BIBdecl

\bibitem{canella2019fallout}
C.~Canella, D.~Genkin, L.~Giner, D.~Gruss, M.~Lipp, M.~Minkin, D.~Moghimi, F.~Piessens, M.~Schwarz, B.~Sunar \emph{et~al.}, ``Fallout: Leaking data on meltdown-resistant {CPU}s,'' in \emph{{ACM SIGSAC}}, 2019, pp. 769--784.

\bibitem{Fredrikson2015ModelInversion}
M.~Fredrikson, S.~Jha, and T.~Ristenpart, ``Model inversion attacks that exploit confidence information and basic countermeasures,'' in \emph{the 22nd {ACM SIGSAC}}, 2015, p. 1322–1333.

\bibitem{9218707}
C.~Gongye, Y.~Fei, and T.~Wahl, ``Reverse-engineering deep neural networks using floating-point timing side-channels,'' in \emph{2020 57th ACM/IEEE Design Automation Conference (DAC)}, 2020, pp. 1--6.

\bibitem{GrussFlushFlush}
D.~Gruss, C.~Maurice, K.~Wagner, and S.~Mangard, ``Flush+flush: A fast and stealthy cache attack,'' in \emph{Detection of Intrusions and Malware, and Vulnerability Assessment}.\hskip 1em plus 0.5em minus 0.4em\relax Springer International Publishing, 2016.

\bibitem{intel2023amx}
\BIBentryALTinterwordspacing
Intel, ``Advanced matrix extensions (amx) for ai acceleration,'' Intel Corporation, 2023. [Online]. Available: \url{https://www.intel.com/content/www/us/en/products/docs/accelerator-engines/advanced-matrix-extensions/ai-solution-brief.html}
\BIBentrySTDinterwordspacing

\bibitem{IntelOptRefManual}
------, ``Intel® 64 and {IA}-32 architectures optimization reference manual,'' \url{https://www.intel.com/content/www/us/en/content-details/814198/intel-64-and-ia-32-architectures-optimization-reference-manual-volume-1.html}, 2024.

\bibitem{kiriansky18dawg}
V.~Kiriansky, I.~A. Lebedev, S.~P. Amarasinghe, S.~Devadas, and J.~S. Emer, ``Dawg: A defense against cache timing attacks in speculative execution processors,'' \emph{MICRO 2018}, pp. 974--987.

\bibitem{kocher2020spectre}
P.~Kocher, J.~Horn, A.~Fogh, D.~Genkin, D.~Gruss, W.~Haas, M.~Hamburg, M.~Lipp, S.~Mangard, T.~Prescher, M.~Schwarz, and Y.~Yarom, ``Spectre attacks: Exploiting speculative execution,'' in \emph{2019 IEEE Symposium on Security and Privacy (SP)}, 2019, pp. 1--19.

\bibitem{kogler2023collide+}
A.~Kogler, J.~Juffinger, L.~Giner, L.~Gerlach, M.~Schwarzl, M.~Schwarz, D.~Gruss, and S.~Mangard, ``$\{$Collide+ Power$\}$: Leaking inaccessible data with software-based power side channels,'' in \emph{32nd USENIX Security Symposium (USENIX Security 23)}, 2023, pp. 7285--7302.

\bibitem{Lipp2021Platypus}
M.~Lipp, A.~Kogler, D.~Oswald, M.~Schwarz, C.~Easdon, C.~Canella, and D.~Gruss, ``{PLATYPUS: Software-based Power Side-Channel Attacks on x86},'' in \emph{2021 IEEE Symposium on Security and Privacy (SP)}.\hskip 1em plus 0.5em minus 0.4em\relax IEEE, 2021.

\bibitem{meltdown}
M.~Lipp, M.~Schwarz, D.~Gruss, T.~Prescher, W.~Haas, J.~Horn, S.~Mangard, P.~Kocher, D.~Genkin, Y.~Yarom \emph{et~al.}, ``Meltdown: Reading kernel memory from user space,'' \emph{Communications of the ACM}, vol.~63, no.~6, pp. 46--56, 2020.

\bibitem{liu2022frequency}
C.~Liu, A.~Chakraborty, N.~Chawla, and N.~Roggel, ``Frequency throttling side-channel attack,'' in \emph{Proceedings of the 2022 ACM SIGSAC Conference on Computer and Communications Security}, 2022, pp. 1977--1991.

\bibitem{RakinDeepSteal2021}
A.~S. Rakin, M.~H.~I. Chowdhuryy, F.~Yao, and D.~Fan, ``Deepsteal: Advanced model extractions leveraging efficient weight stealing in memories,'' in \emph{2022 IEEE Symposium on Security and Privacy (SP)}, 2022, pp. 1157--1174.

\bibitem{schwarz2019zombieload}
M.~Schwarz, M.~Lipp, D.~Moghimi, J.~Bulck, J.~Stecklina, T.~Prescher, and D.~Gruss, ``\BIBforeignlanguage{English}{Zombieload: Cross-privilege-boundary data sampling},'' in \emph{\BIBforeignlanguage{English}{CCS 2019}}, Nov. 2019, pp. 753--768.

\bibitem{TramerZJRR16}
F.~Tram\`{e}r, F.~Zhang, A.~Juels, M.~K. Reiter, and T.~Ristenpart, ``Stealing machine learning models via prediction apis,'' in \emph{Proceedings of the 25th USENIX Conference on Security Symposium}, 2016, p. 601–618.

\bibitem{tramèr2019slalomfastverifiableprivate}
\BIBentryALTinterwordspacing
F.~Tramèr and D.~Boneh, ``Slalom: Fast, verifiable and private execution of neural networks in trusted hardware,'' 2019. [Online]. Available: \url{https://arxiv.org/abs/1806.03287}
\BIBentrySTDinterwordspacing

\bibitem{wang2022hertzbleed}
Y.~Wang, R.~Paccagnella, E.~T. He, H.~Shacham, C.~W. Fletcher, and D.~Kohlbrenner, ``Hertzbleed: Turning power {Side-Channel} attacks into remote timing attacks on x86,'' in \emph{USENIX Security 22}, pp. 679--697.

\bibitem{wei2018know}
L.~Wei, B.~Luo, Y.~Li, Y.~Liu, and Q.~Xu, ``I know what you see: Power side-channel attack on convolutional neural network accelerators,'' in \emph{Proceedings of the 34th Annual Computer Security Applications Conference}, 2018, pp. 393--406.

\bibitem{yan2018invisispec}
M.~Yan, J.~Choi, D.~Skarlatos, A.~Morrison, C.~Fletcher, and J.~Torrellas, ``Invisispec: Making speculative execution invisible in the cache hierarchy,'' in \emph{2018 MICRO}.\hskip 1em plus 0.5em minus 0.4em\relax IEEE, 2018, pp. 428--441.

\bibitem{Yan2020Cachetelepathy}
M.~Yan, C.~W. Fletcher, and J.~Torrellas, ``Cache telepathy: Leveraging shared resource attacks to learn {DNN} architectures,'' in \emph{29th USENIX Security Symposium (USENIX Security 20)}, Aug. 2020, pp. 2003--2020.

\end{thebibliography}
\end{spacing}
\vfill
\end{document}